\documentstyle[aps,pra]{revtex}
\title {Grover's search algorithm and the quantum measurement problem}
\author{Manoj K. Samal and Partha Ghose} \address{S. N. Bose National
Centre for Basic Sciences, JD/III, Salt Lake, Kolkata 700 098, India}
\newcommand{\be}{\begin{equation}}
\newcommand{\ee}{\end{equation}}
\newcommand{\ben}{\begin{eqnarray}}
\newcommand{\een}{\end{eqnarray}}
\newcommand{\bF}{\begin{figure}}
\newcommand{\eF}{\end{figure}}
\begin{document}
\maketitle

\begin{abstract}
It is suggested that the individual outcomes of a measurement process can
be understood within standard quantum mechanics in terms of the measuring
apparatus, treated as a quantum computer, executing Grover's search
algorithm.
\end{abstract}

\section{Introduction}

The measurement problem lies at the root of all interpretational
problems of quantum mechanics\cite{wz}. Once the wavefunction of a
system is assumed to contain the most complete information
possible of the system, the measurement problem is inevitable. Let
a system be in the pure state
\be
|S\rangle = \sum_i^n c_i |S_i\rangle, \,\,\,\,\, \sum_i^n |c_i|^2
= 1\ee where $|S_i\rangle$ are a complete set of eigenstates of
some observable $A$. If one measures $A$ on this state, one would
get the result $a_i$ with probability $|c_i|^2$, and once the
measurement is complete, the state is forced into the eigenstate
$|S_i\rangle$. This does not imply that the system is a
statistical ensemble of these states $|S_i\rangle$ with
probabilities $|c_i|^2$ in the sense of classical probability
theory, and the measurement simply removes the ignorance. The
simplest way to see this is to compare the density matrix of a
pure state $|\Psi\rangle$,
\be
(\rho)_{ij} = (|\Psi\rangle\langle \Psi|)_{ij}= c_i^*
c_j\label{1}\ee which is non-diagonal with that for a `mixture',
\be
(\hat{\rho})_{ij} = |c_i|^2 \delta_{ij}\label{2}\ee which is
diagonal. Although (\ref{1}) and (\ref{2}) give identical results
for the probabilities of obtaining the various eigenvalues $a_i$
of the observable $A$, they predict quite different results for
observables that do not commute with $A$.

According to standard measurement theory a measurement quenches
the off-diagonal interference terms and reduces the density matrix
of a pure state like (\ref{1}) to the diagonal form (\ref{2}).
This loss of coherence or reduction of the state vector cannot
result from Schr\"odinger evolution which is unitary, causal and
reversible. A measurement interaction first entangles a system $S$
with the measuring apparatus $X$. In general one obtains the state
\be
|\Psi\rangle = \sum_i^n c_i |S_i\rangle |X_i\rangle \label{mpr}
\ee

\noindent where the states $|X_i\rangle$ span the pointer basis.
This is a unitary Schr\"odinger process that von Neumann calls
`process 2'\cite{vn}. It correlates every state $|S_i\rangle$ with
a definite apparatus state $|X_i\rangle$. Since, however, this is
an entangled state, it has to be reduced to a particular state
$|S_i\rangle |X_i\rangle$ before the result can be read off.
`Process 1' is a non-unitary process that achieves this by
projecting the state $|\Psi\rangle$ to this state with the help of
the projection operator $\Pi_i = |X_i\rangle \langle X_i|$. One
then obtains the reduced density matrix
\be
\rho \rightarrow \hat{\rho} = \sum_i \Pi_i \rho \Pi_i \ee which is
diagonal and represents a heterogeneous mixture with probabilities
$|c_i|^2$. This is the least understood aspect of quantum
mechanics and lies at the root of all its interpretations.

There is no universally accepted solution to this problem within
standard quantum mechanics. The many-worlds interpretation tries
to solve this problem by rejecting the projection postulate and
postulating instead that the universe splits into $n$ orthogonal
universes at every measurement, each carrying one possible result
of the measurement, and that no communication is possible between
these universes\cite{everett}. This is considered by some as the
only possible solution to the problem, but is rejected by many on
aesthetic grounds (it being extravagant in its unverifiable
profusion of parallel universes). Environment induced
decoherence\cite{deco} was also thought to be an alternative
solution, but it has become clear now that although an extremely
useful concept of great practical importance in its own right, it
does not actually solve the measurement problem as it has no
explanation for the occurrence of individual events, i.e., why all
diagonal elements of the reduced density matrix except one vanish
for a single process (like the blackening of a single spot on a
photographic plate). It is in this context that the Grover search
algorithm\cite{grover1} offers a plausible solution.

\section{The Grover search algorithm}

Suppose one wants to search a telephone number in a telephone
directory of a large city with $N$ entries. A classical computer
will have to carry out $\cal{O}(N)$ operations. Grover's algorithm
can speed up this search and complete it in $\cal{O}(\sqrt{N})$
operations on a quantum computer. We wish to point out that this
search algorithm also offers a plausible solution to the
measurement problem. The essential point is that the algorithm
which is essentially of `process 2' type amplifies the amplitude
of an identified target (the amplitude corresponding to a
particular eigenstate in this case) at the cost of all other
amplitudes to a point where the latter become so small that they
cannot be recorded by detectors of finite efficiency. This
therefore replaces von Neumann's `process 1'.

Let us see how this can happen. Let the set \{$|S_i\rangle
|X_i\rangle$\} (where i=1, 2, ..., N) in eqn. (\ref{mpr}) be the
search elements that a quantum computer inside the apparatus has
to deal with. Let these elements be indexed from $0$ to $N - 1$.
This index can be stored in $n$ bits where $N = 2^n$. Let the
search problem have exactly $M$ solutions with $1 \leq M \leq N$.
Let $f(\xi)$ be a function with $\xi$ an integer in the range $0$
to $N -1$. By definition $f(\xi) = 1$ if $\xi$ is a solution to
the search problem and $f(\xi) = 0$ if $\xi$ is not a solution to
the search problem. One then needs an {\it oracle} that is able to
recognize solutions to the search problem\cite{nielsen}. This is
signalled by making use of a {\it qubit}. The oracle is a unitary
operator $O$ defined by its action on the computational basis:
\be
O:\,\,\,\,|\xi\rangle |q\rangle \rightarrow |\xi\rangle | q \oplus
f(\xi)\rangle\ee where $|\xi\rangle$ is the index register,
$\oplus$ denotes addition modulo $2$, and the oracle qubit
$|q\rangle$ is a single qubit that is flipped if $f(\xi) = 1$ and
is unchanged otherwise. Thus,

\ben |\xi\rangle |0\rangle &\rightarrow&  |\xi\rangle |0\rangle
\,\,\,\,{\rm if}\,\,|\xi\rangle\,\, {\rm is\,\, not\,\, a\,\,
solution}\\|\xi\rangle |0\rangle &\rightarrow& |\xi\rangle
|1\rangle\,\,\,\,{\rm if}\,\,|\xi\rangle\,\, {\rm is\,\, a\,\,
solution}\een It is convenient to to apply the oracle with the
oracle qubit initially in the state $|q\rangle =(|0\rangle -
|1\rangle)/\sqrt{2}$ so that

\be
O:\,\,\,\,|\xi\rangle |q\rangle \rightarrow (- 1)^{f(\xi)}
|\xi\rangle |q\rangle\ee Then the oracle marks the solutions to
the search by shifting the phase of the solution. If there are $M$
solutions, it turns out that one need only apply the search oracle
$\cal{O}(\sqrt{N/M})$ times on a quantum computer.

To start with, the quantum computer, assumed to be an integral
part of the final detector, is always in the state
$|0\rangle^{\otimes n}$. The first step in the Grover search
algorithm is to apply a Hadamard transform to put the computer in
the equal superposition state
\be
|\psi\rangle = \frac{1}{\sqrt{N}} \sum_{\xi=0}^{N - 1}
|\xi\rangle\label{eq}\ee The search algorithm then consists of
repeated applications of the Grover iteration or Grover operator
$G$ which can be broken up into the following four operations:
\begin{enumerate}
\item The oracle $O$.
\item The Hadamard transform $H^{\otimes n}$.
\item A conditional phase shift on the computer with every
computational basis state except $|0\rangle$ receiving a phase
shift of $- 1$, i.e., $$|\xi\rangle \rightarrow (- 1)^{f (
\xi)}|\xi\rangle$$
\item The Hadamard transform $H^{\otimes n}$.
\end{enumerate}
These operations can be carried out on a quantum computer. The
combined effect of steps $2,3$ and $4$ is
\be
G = H^{\otimes n} (2 |0\rangle \langle 0| - I)H^{\otimes n} = 2
|\psi\rangle \langle \psi| - I\ee where $|\psi\rangle$ is given by
(\ref{eq}).

The Grover operator $G$ can be regarded as a rotation in the two
dimensional space spanned by the vector $|\psi\rangle$ which is a
uniform superposition of the solutions to the search problem. To
see this, define the normalized states

\ben |\alpha\rangle &=& \frac{1}{\sqrt{N - M}}\sum_\xi^{''}
|\xi\rangle\\ |\beta\rangle &=& \frac{1}{\sqrt{M}}\sum_\xi^{'}
|\xi\rangle\een where $\sum_\xi^{'}$ indicates a sum over all
$\xi$ that are solutions to the search problem and $\sum_\xi^{''}$
a sum over all $\xi$ that are not solutions to the search problem.
The the initial state can be written as \be |\psi\rangle =
\sqrt{\frac{N - M }{N}}|\alpha\rangle +
\sqrt{\frac{M}{N}}|\beta\rangle\ee so that the apparatus (with the
built-in quantum computer) is in the space spanned by
$|\alpha\rangle$ and $|\beta\rangle$ to start with. Now notice
that the oracle operator performs a rotation about the vector
$|\alpha\rangle$ in the plane defined by $|\alpha\rangle$ and
$|\beta\rangle$, i.e.,
\be
O (a |\alpha\rangle + b |\beta\rangle) =  a |\alpha\rangle - b
|\beta\rangle\ee Similarly, $G$ also performs a reflection in the
same plane about the vector $|\psi\rangle$, and the effect of
these two reflections is a rotation. Therefore, the state $G^k
|\psi\rangle$ remains in the plane spanned by $|\alpha\rangle$ and
$|\beta\rangle$ for all $k$. The rotation angle can be found as
follows. Let ${\rm cos} \theta/2 =\sqrt{N - M/N}$ so that
$|\psi\rangle = {\rm cos} (\theta/2) |\alpha\rangle + {\rm
sin}(\theta/2)|\beta\rangle$. Then one can show that\cite{nielsen}
\be
G |\psi\rangle = {\rm cos}\frac{3 \theta}{2}|\alpha\rangle + {\rm
sin} \frac{3 \theta}{2}|\beta\rangle\ee so that $\theta$ is indeed
the rotation angle, and so

\be G^k |\psi\rangle = {\rm cos}(\frac{2k
+1}{2}\theta)|\alpha\rangle + {\rm sin}(
\frac{2k+1}{2}\theta)|\beta\rangle\ee Thus, repeated applications
of the Grover operator rotates the vector $|\psi\rangle$ close to
$|\beta\rangle$. When this happens, an observation in the
computational basis produces one of the outcomes superposed in
$|\beta\rangle$ with high probability. Grover has shown later that
the Hadamard transforms can be replaced by more general transforms
\cite{grover2}. Finally, we would like to point out that in a
quantum measurement only one outcome must occur and hence, the
number $M$ of simultaneous solutions that the Grover algorithm
searches is always unity.

Let us see how it works out in the typical case of the
Stern-Gerlach experiment. The index register in this case consists
of the two states $|X_\uparrow\rangle |\uparrow\rangle $ and
$|X_\downarrow\rangle |\downarrow\rangle$ where
$|X_\uparrow\rangle = |(x_1,y_1)\rangle$ and $|X_\downarrow\rangle
= |(x_2,y_2)\rangle$ correspond to the two spots on a
two-dimensional screen in the $(x, y)$ plane. Since $M = 1$ and
$N=2$, the initial state of the detector is a $50-50$
superposition of $|\alpha\rangle$ and $|\beta\rangle$. What
happens in an individual event (the appearance of a single spot at
certain times either at $(x_1,y_1)$ or $(x_2,y_2)$) is that the
Grover search $G$ rotates the initial state of the apparatus to
one of the two solutions, each solution occurring with $50\%$
probability.

\section{Conclusions}

What we have shown is that the Grover search is a plausible
explanation of how individual events occur in standard quantum
mechanics. The entire process is unitary, i.e., of type 2 but yet
it is possible to get information from these events because the
amplitudes that are made arbitrarily small by the Grover search
cannot be detected by measuring devices of finite efficiency. The
projection postulate accounts for individual events but at the
expense of unitarity. The decoherence approach, on the other hand,
preserves unitarity but fails to account for individual events. In
this sense the explanation in terms of the Grover search is
different from both the von Neumann projection postulate and the
environmentally induced decoherence approach. This proposal is
also entirely different from the many-worlds interpretation as it
works in a single universe. The mystery of the von Neumann
projection or the Everett splitting of the universe is replaced by
the riddle of how precisely a measuring apparatus executes the
Grover algorithm.

\section{Acknowledgement}

We would like to thank the Tata Institute of Fundamental Research,
Mumbai for their hospitality and providing a stimulating
environment during the recently held school on Quantum Physics and
Information Processing (18-27 March, 2002) during which we
mentioned to Lov Grover the possibility of using the quantum
search algorithm as a solution to the measurement problem. We are
also thankful to 1998 Computing Ltd., UK for financial support to
attend the Workshop and to the Department of Science and
Technology, Government of India for a research grant that enabled
this work to be undertaken.

\end{document}